\documentclass{aa}  

\usepackage{graphicx}
\usepackage{txfonts}
\usepackage{color}

\begin{document}

     \title{Constraining the thin disc initial mass function using Galactic classical Cepheids}

%   \subtitle{}

\author{
R. \,Mor\inst{1} 
\and A.C. \, Robin\inst{2}
\and F. \, Figueras\inst{1}
\and B. \, Lemasle\inst{3}
          }

 \institute{
Dept. F\'isica Qu\`antica i Astrof\'isica, Institut de Ci\`encies del Cosmos, Universitat de Barcelona (IEEC-UB), Mart\'i Franqu\`es 1, E08028 Barcelona, Spain.
\email{rmor@fqa.ub.edu}
\and
Institut Utinam, CNRS UMR6213, Universit\'e de Bourgogne Franche-Comt\'e, OSU THETA , Observatoire de Besan\c{c}on, BP 1615, 25010 Besan\c{c}on Cedex, France
\and
Astronomisches Rechen-Institut, Zentrum für Astronomie der Universität Heidelberg, Mönchhofstr. 12-14, D-69120 Heidelberg, Germany
}

   \date{Received August 3, 2016; accepted October 20, 2016}

% \abstract{}{}{}{}{} 
% 5 {} token are mandatory
 
  \abstract
   {The initial mass function (IMF) plays a crucial role in galaxy evolution and its implications on star formation theory make it a milestone for the next decade. It is in the intermediate and high mass ranges where the uncertainties of the IMF are larger. This is a major subject of debate and analysis both for Galactic and extragalactic science.} 
   { Our goal is to constrain the IMF of the Galactic thin disc population using both Galactic classical Cepheids and Tycho-2 data.}
   { For the first time, the Besan\c{c}on Galaxy Model (BGM) has been used to characterize the Galactic population of  classical Cepheids. We  modified the age configuration in the youngest populations of the BGM thin disc model to avoid artificial discontinuities in the age distribution of the simulated Cepheids. Three statistical methods, optimized for different mass ranges, have been developed and applied to search for the best IMF that fits the observations. This strategy enables us to quantify variations in the star formation history (SFH), the stellar density at Sun position and the thin disc radial scale length. A rigorous treatment of unresolved multiple stellar systems has been undertaken, adopting a spatial resolution according to the catalogues used.  }
      %  results heading (mandatory)
   {For intermediate masses, our study favours a composite field-star IMF slope of $\alpha=3.2$ for the local thin disc, excluding flatter values, e.g. the Salpeter IMF ($\alpha=2.35$). Our findings are broadly consistent with previous results derived from Milky Way models.
Moreover, a constant SFH is definitively excluded, the three statistical methods considered here show that it is inconsistent with the observational data. }     
  {Using field stars and Galactic Classical Cepheids, we  found   an IMF steeper than the canonical stellar IMF of associations and young clusters above
$1M_\odot$. This result is consistent with the predictions of the integrated Galactic IMF.}

\titlerunning{Constraining the IMF using Galactic Cepheids}
   \maketitle
%
%________________________________________________________________

\section{Introduction}

Classical Cepheids are probably the best-known and most important pulsating variable stars. Since Henrietta Swan Leavitt determined for the first time, in 1912, their period-luminosity relation \citep{Leavitt1912}, classical Cepheids have become the first ladder of the extragalactic distance scale, since they provide accurate distances in the Local Universe. Now, in the Gaia era, the expected thousands of Cepheids that are going to be detected \citep{Eyer2000,Windmark2011,Mor2015}, again place  these classical variable stars in a privileged position when studying the structure of the Milky Way. In this work we plan to update the youngest populations in the Besan\c{c}on Galaxy Model (BGM) making it capable to simulate a reliable population of Cepheids. With this new configuration, and as a first step, we aim to use the classical Cepheids as tracers of intermediate-mass population to constrain the initial mass function (IMF) of the thin disc.

The IMF describes the mass distribution of a star formation episode.  Together with the star formation history (SFH), it is one of the most important functions to characterise the evolution of the stellar populations in the Milky Way and external galaxies. The chemical composition and luminosity of galaxies is directly influenced by the IMF as it determines the baryonic content and the light of the Universe.  \cite{Salpeter1955} was the first to describe the IMF as a simple power law $dN= \xi (m)dm=k  m^{-\alpha}dm$ and he estimated a power-law index of $\alpha=2.35,$ taking into consideration an age of
6 Gyr for the Milky Way. Since then, several fundamental works on the  empirical derivation of the Galactic IMF have been written, such those of \cite{Schmidt1959}, \cite{MS79},  \cite{Scalo1986}, \cite{KTG93}, and \cite{KroupaSci2002} among others. Even so, the shape of the IMF is still a matter of debate, in particular for the high and intermediate stellar mass range.

Several methods to derive the IMF consist either in analysing the mass distribution of complex preprocessed samples, or by fitting models to star counts in complete but limited samples. Advantages and drawbacks of this last method are discussed in this paper, keeping in mind that this approach  depends on other parameters such as the SFH, the density distribution, or the interstellar extinction. 

An important handicap when studying the IMF at intermediate and high masses is the low number of stars of these masses that can be found in clusters or associations. The same occurs when using field stars just around few hundred parsecs from the Sun. Classical Cepheids can solve the problem of poor statistics at intermediate masses because they are bright and their distance can be accurately determined. Soon Gaia will provide us with thousands of them, but for the moment the most complete catalogues of Galactic classical Cepheids provide us with hundreds of Cepheids in the intermediate stellar mass range. These hundreds of Cepheids, together with the $\approx 800000$ stars from the Tycho-2 catalogue \citep{Hog2000} up to $V=11$ (that is dominated by  low-mass stars), ensures good statistics to constrain the IMF in the present work.

It have been suggested that the IMF is closely related to the structure and fragmentation mechanism of  molecular clouds where the stars are formed. Thus, several attempts to derive the IMF theoretically have been made in this context. For example, \cite{Adams1996} computed a semi-empirical mass formula (SEMF) which provides the transformation between initial conditions in molecular clouds and the final masses of forming stars based on the idea that stars determine their own masses through the action of powerful stellar outflows. For a particular SEMF, a given distribution of initial conditions theoretically predicts a corresponding IMF. Key works when discussing the theory of the IMF are the following: \cite{Larson1992}, assuming a two-dimensional molecular cloud and hierarchical fragmentation, derived a slope of $\alpha=3$ for high masses and $\alpha=3.3$ for a more general IMF. In contrast, \cite{Padoan2002}, considering a turbulent fragmentation of  molecular clouds, derived an IMF with a slope $\alpha=2.33$, significantly lower and very close to the Salpeter one. \cite{Elmegreen1997} obtained a slope value of $\alpha \approx 2-2.7$ considering a turbulent
fractal molecular cloud. We need to bear in mind that several parameters are degenerated when deriving the theoretical formation of the stellar cores from the molecular clouds, i.e. coalescence of protostellar cores, mass-dependence, accretion process, stellar feedback, or fragmentation. In the case of intermediate and high-mass stars the formation process is even more complex. Thus, in this context, the empirical and semi-empirical determination of the IMF at intermediate and high masses can contribute to the understanding of the star formation mechanism.

From a population synthesis point of view, several attempts have made  to determine the IMF for different components of the Milky Way. For example in \cite{Reyle2001} and \cite{Robin2014}, the IMF of the thick disc component was studied using deep star counts in different directions. Moreover in \cite{Robin2000}, the halo IMF was  investigated. Recently the IMF of the thin disc was evaluated by our team (\citealt{Czekaj2014}) using Tycho-2 data and also by \cite{Rybizki2015} using an observational sample that consists of stars from the extended Hipparcos catalogue and the Catalogue of Nearby Stars.

In this work, we aim to constrain the thin disc IMF at intermediate masses using the BGM. This tool is being updated each year by studying the different components of the Milky Way, e.g. the bulge region \citep{Robin2012}, where a triaxial boxy shape for the bar is fitted; the thick disc, \cite{Robin2014}, where two successive star formation episodes are proposed; the thin disc \citep{Czekaj2014}, where results point to a decreasing SFH whatever IMF is assumed. Furthermore the BGM has been used to study the interstellar medium \citep{Marshall2006}, Galactic kinematics and dynamics (e.g. \citealt{Bienayme2015}), to estimate micro-lensing probabilities  \citep{Awiphan2016,Kerins2009} and BGM has also been  used for the preparation of the ESA Gaia astrometric mission \citep{GUMS2012}. In a daily effort to update the BGM piece by piece, contributing step by step to the knowledge of the different components of the Milky Way, in this work we aim to update the youngest populations of the BGM thin disc, improving its fit with the Tycho-2 data and using it to constrain the IMF.

In Section 2, we briefly describe the BGM, the model ingredients, the characteristics of  classical Cepheids in the BGM and our strategy. In Section 3, we describe the observational sample. In Section 4, we present our evaluation methods. Results are presented in Section 5, while discussion and conclusions are presented in Sections 6 and 7.

\section{The Besan\c{c}on Galaxy Model}\label{BGM}

The Besan\c{c}on Galaxy Model has  proved to be an efficient tool to test the Milky Way galaxy structure and evolution scenarios.  Last updates are described in \cite{Robin2003}, \cite{Robin2012}, \cite{Robin2014} and \cite{Czekaj2014}, whereas its dynamical consistency  is discussed in \cite{Bienayme1987} and \cite{Bienayme2015}. In this study, we are interested in generating a full sky set of data to be compared with both  Tycho-2 data and the most complete catalogues of Galactic Cepheids. These catalogues are complete up to a bright limit in apparent magnitude, thus with a dominant contribution from the thin disc population and a small contribution from the thick disc (expected $\approx 10\%$ ) and the halo (expected $\approx 0.3\%$).  

\subsection{The thin disc population}\label{ThinDisc}

The thin disc component is described in \cite{Czekaj2014}. The stars  are generated following an IMF and an SFH, with a continuous star formation during the disc evolution. The thin disc population is divided into seven age sub-populations with the age intervals described in \cite{Bienayme1987}. The density distribution of each sub-population of the thin disc is assumed to follow an Einasto density profile as described in \cite{Robin2012} Section 2.1, except for the youngest sub-population which follows the expression described in \cite{Robin2003}. The main parameters for the characterization of the Einasto profiles are the eccentricities of the ellipsoid ($\epsilon$, that is the axis ratio), the scale length of the disc ($R_d$) and the scale length of the disc hole ($R_h$). A velocity dispersion as a function of age is adopted and, each time the IMF and the SFH is changed, new structure parameters (e.g. the eccentricities of the ellipsoids $\epsilon$) are computed to keep the dynamical statistical equilibrium \citep{Bienayme1987}. Stellar evolutionary tracks and model atmosphere, combined with an age-metallicity relation, enable us to go from masses, ages, and metallicities to the space of the observables. In this process a 3D interstellar extinction model is assumed. 

BGM thin disc simulations work following the scheme of Figure 3 from \cite{Czekaj2014}. The SFH, a key ingredient of the simulation, determines the amount of stars generated in each one of the seven age sub-populations. Once a star is created, we assign an age and a metallicity to it. The ages are drawn randomly from the uniform distribution in the interval of the given age sub-population. The metallicity is drawn for each star from its own age, according to the age-metallicity relation adopted. When the age, mass, and metallicity are established, we interpolate the stellar evolutionary tracks and find the position of the star in the Hertzsprung-Russell diagram.

The model includes the generation of unresolved and resolved binary systems according to an imposed spatial resolution. The binarity treatment is well described in Section 2.2.2 and 4.3. of \cite{Czekaj2014}. Binaries are generated following the scheme proposed by \cite{Arenou2011}, also used to generate the Gaia Universe Model \citep{GUMS2012}. To decide if each newly created star is either a single or primary component of multiple system, the model uses a probability density function that depends on the mass of the object and its luminosity class. The mass-ratio distribution between system components, estimated from observations \citep{Arenou2011}, takes into account the spectral type and mass of the primary component.

Hereafter, our initial default model (hereafter DM) will be Model B of \cite{Czekaj2014}. Model B is the model proposed in \cite{Czekaj2014}, which gave better results in different studies (e.g. \citealt{Robin2014}), moreover it is used nowadays as the thin disc model for the Gaia Object Generator \citep{Robin2012,Luri2014}. Table 5 of \cite{Czekaj2014} shows the set of thin disc ingredients adopted to fit Tycho-2 data using the  radial scale length parameters detailed in \cite{Robin2003}. The stellar evolutionary models used are those of \cite{Chabrier1997} for $M < 0.7 M_\odot$, \cite{Bertelli2008,Bertelli2009} for $0.7 M_\odot < M < 20 M_\odot,$ and 
\cite{Bertelli1994} $20 M_\odot < 120 M_\odot$.

\subsection{Generation of classical Cepheids}\label{CphBGM}

The instability strip (IS) is the region of the Hertzprung-Russel diagram occupied by the pulsating variable stars, including classical Cepheids. The hotter and cooler boundaries of the IS are called the blue edge and red edge, respectively. For this work, up to apparent magnitude $V=9$, that is for stars around solar neighbourhood, we adopt the blue edge as $log(T_{eff}) = -( log(\mathcal{L}/\mathcal{L_\odot})-62.7 )/15.8 $ from \cite{Bono2000}  and red edge as $log(T_{eff}) = -(log( \mathcal{L}/\mathcal{L_\odot})-40.2)/10.0$ from \cite{Fiorentino2013}, both derived from Cepheid pulsation models at solar metallicity. For Cepheids at larger heliocentric distances (magnitudes $9<V \leq 12$), and given the radial metallicity gradient of the Milky Way (e.g. \citealt{Genovali2014}), we  decided to keep the same red age and to use as the blue edge the one derived from pulsation models at lower metallicity ($z=0.008$) from \cite{Fiorentino2013}, that is $log(T_{eff}) = -(log( \mathcal{L}/\mathcal{L_\odot})-52.5 )/13.1$.

We impose a luminosity cut in the range $2.7 \leq log(\mathcal{L}/\mathcal{L_\odot}) \leq 4.7$. This  luminosity cut constrains the  effective temperature of the Cepheids  to about $4000 \leq T_{eff} \leq 7000K$ \citep{Bono1999}.

\begin{figure}
   \centering
   \includegraphics[width=\hsize]{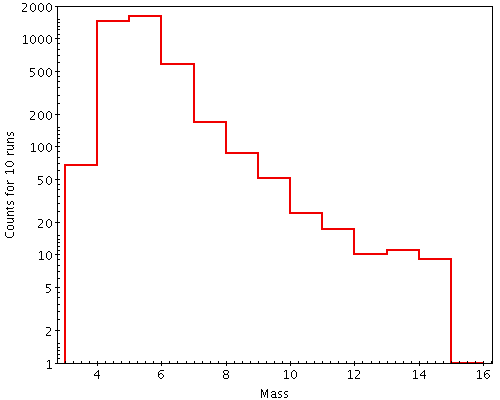} 
      \caption{Mass distribution of the simulated Galactic classical Cepheids for 10 runs. The plot corresponds to the default model. The mass is expressed in solar masses.}
         \label{massrange}
\end{figure}

The masses of our simulated Cepheids, selected with the boundaries of the IS described above, are found to be between 3 and 11 $M_\odot$  with a few reaching up to 15 $M_\odot$ (see Figure \ref{massrange}). These values are in good agreement with the mass ranges in the literature .  To start with, the classical review  from \cite{Cox1980} established the  Cepheid mass range between $3$ and $15 M_\odot$. More recent studies, such as \cite{Caputo2005}, found masses between $5$ to $15 M_\odot$. Using evolutionary models, \cite{Bono2000b} estimated a minimum mass of $\approx 3.25M_\odot$ for low metallicity Cepheids and $\approx 4.75 M_\odot$ for Cepheids with solar metallicity. The upper limit given by both \cite{Bono2000b} and \cite{Alibert1999} could depend on metallicity and it is in the range $10-12 M_\odot$. More recently, \cite{Anderson2014}, also using stellar model but including stellar rotation, predict masses from $4$ to $10 M_\odot$. This agreement between literature and our generated Cepheid mass distribution using the BGM reinforce both the boundaries of the IS adopted and the BGM Cepheid-generation strategy. 

The age distribution of our simulated Cepheids is presented in Figure \ref{AgeGap}.  Most of our simulated Cepheids have ages of between $20$ and $200 Myrs$, well compatible with the values derived by \cite{Bono2005}, who estimated ages  from $\approx 25 Myrs$ to $\approx 200 Myrs$, based on evolutionary and pulsation models. The shape of the age distribution obtained here using BGM is discussed in Section \ref{AgeGapSec}.

With regard to binary fraction, our simulations show that about 68\% of  classical Cepheids are contained in a binary system. This percentage is consistent with values from the literature (e.g. \citealt{Szabados2003} estimated about 60-80\% of Cepheids). None of our simulated Cepheids are secondary components of a multiple system.  To generate these objects as secondaries, primary stars should have  $M>3.5M_\odot$ and we have checked that only about 2\% of stars up to $V=12$ fulfil this condition. Furthermore, this probability becomes negligible when imposing the secondary to be in the instability strip.
 
\cite{Czekaj2014} adopted a spatial resolution of 0.8 arcsec according to the resolution of Tycho-2 catalogue. In the cited work, it was noted that most of the simulated binaries have angular separation smaller than 0.5 arcsec. The Cepheid catalogues used here could have worse resolutions than Tycho-2, thus the distribution of the simulated angular separation enables us to adopt the same 0.8 arcsec resolution for the Cepheids without compromising the total number of star counts.

\subsection{Cepheid ages to constrain the BGM youngest populations}\label{AgeGapSec}

\begin{figure} 
   \centering
   \includegraphics[width=\hsize]{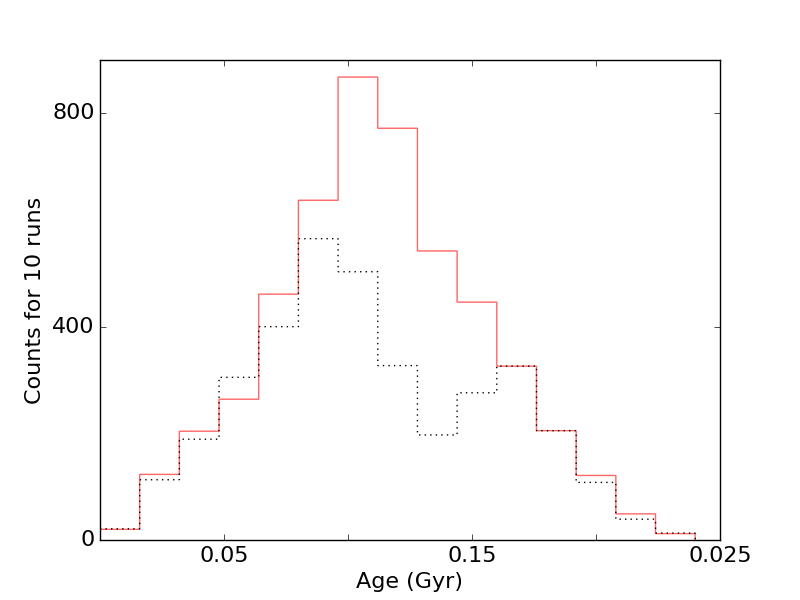}
      \caption{Age distribution of the simulated Galactic classical Cepheids for 10 runs. The black dotted line is for the initial default model. The red solid line is for the modified default model after the update of the age intervals of the youngest populations of the BGM.}
         \label{AgeGap}
\end{figure}

In Figure \ref{AgeGap}, we show the age distribution of the simulated classical Cepheids with a black dotted line  up to apparent magnitude $V=12$ using the DM. Like the other stars, the Cepheids are generated following an SFH and an IMF, as described in Section 2.1. Since classical Cepheids are young stars they belong to the sub-population 1 and 2 of the thin disc component of the BGM. Although the range of the ages shown in Figure \ref{AgeGap} is compatible with the ages of Cepheids \citep{Bono2005}, its distribution presents a double peak which has not been seen in the empirical data. This is an artefact that comes from a discontinuity between  sub-populations 1 and 2 of the thin disc in the BGM. In the present work, we  modified the age interval of these two youngest sub-populations to avoid this discontinuity. In the initial DM, the first population covered an age interval between 0 and 0.15 Gyr, while the second sub-population covered the age range 0.15 to 1 Gyr. From now on, the age range of the first sub-population and second sub-population of our DM will be set from 0.0 to 0.10 Gyrs and between 0.10 to 1 Gyrs respectively. Thus, once the age interval for sub-population 1 and 2 is redefined, all the machinery of the BGM is set up and rerun, e.g. the amount of stars generated in the so-called new sub-population 1 and 2 is derived from the SFH, and the integrals over time will be computed according to the new age range. A new BGM simulation is used to update the Cepheid age distribution. In Figure \ref{AgeGap}, we show in red the age distribution of the simulated Galactic classical Cepheids with the revised DM. Now the age distribution follows a smooth single peak distribution. As seen in  Section \ref{results}, this modification of the age ranges  improves the fit of the BGM with Tycho-2 data in the Galactic Plane. 

\subsection{Model variants and strategy}\label{mv}

\begin{figure*} 
   \centering
   \includegraphics[width=\hsize]{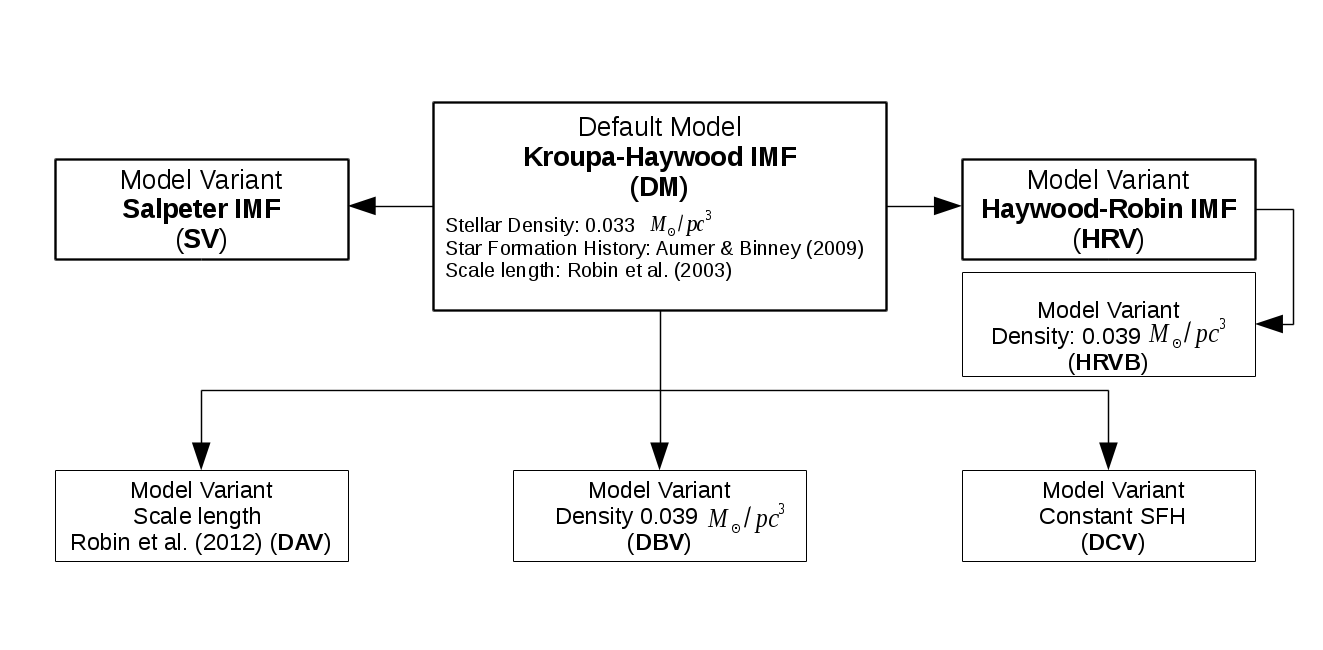}
      \caption{Scheme of the seven model variants tested in the present paper. For the three main variants, DM, HRV, and SV only the IMF has been changed.  The DCV differs from the DM in the SFH, DBV differs from DM in the stellar density at Sun position, and DAV differs from DM on the thin disc radial scale length. The HRV has its own variant in stellar density at Sun position, the HRVB.}
         \label{variants}
   \end{figure*}

In Figure \ref{variants}, we present the scheme of the seven model variants proposed to constrain the thin disc IMF. To evaluate which is the best of the tested IMFs (see Section \ref{IMFs}) it is mandatory to analyse not only the changes that are due to the IMF, but also evaluate the impact of other key ingredients on the mean properties of the simulated catalogues. Work performed in \cite{Czekaj2014}, \cite{Robin2012}, and previous experience has allowed us to identify the SFH, the stellar density at the Sun position, and the radial scale length as the key parameters most affecting  the star count analysis that we perform in this paper.

By departing from the DM, changing only the IMF, we  constructed the two other  main model variants (see Fig. \ref{variants}). Those are the Salpeter Variant (SV) and the Haywood-Robin variant (HRV). In Section \ref{IMFs}, we further describe the  IMFs used. The impact of thin disc radial scale length variations is tested, assuming the scale length of \cite{Robin2003} ($R_d=2530pc$ and $R_h=1320pc$) for the DM  and changing it to the values of \cite{Robin2012} ($R_d=2170pc$ and $R_h=1330pc$) in the default model A variant (DAV). These new scale lengths have been obtained by fitting 2MASS data towards the bulge region in \cite{Robin2012}. To evaluate the effects of the variation in the local stellar mass density, we  tested the values of \cite{Wielen1974} ($0.039 M_\odot/pc^3$ ) in the default model B variant (DBV) and \cite{JahreiB1997} ($0.033 M_\odot/pc^3$ ) used in the DM. Both values are selected because they are used in the best-fit models in \cite{Czekaj2014} and they represent a good range of the values published in the literature. To analyse the effects produced by changes in the SFH, we  assumed the decreasing SFH by \cite{Aumer&Binney2009} for the DM and a constant SFH in the default model C variant (DCV). 
Additionally, to further study the Haywood-Robin IMF, we test the variant HRVB, which has the HRV parameters but using $0.039 M_\odot/pc^3$ as the local stellar mass density.

For the sky areas with longitude between -100 and 100 and latitudes between -10 and 10 the model variants presented in Figure \ref{variants} have been tested with two different extinction models, \cite{Marshall2006} and \cite{Drimmel2001}. Using this strategy, we are able to identify what  the impact of the interstellar extinction is in our Cepheids data, most of them contained in the Galactic plane.  All the other thin disc model ingredients are maintained as fixed. Its values are those adopted by \cite{Czekaj2014}, Tables 2 and 5. Once a full set of parameters is adopted, thus for each of the seven variant models in Figure \ref{variants}, we  applied dynamical constraints as described in \cite{Bienayme1987}, solving the Poisson equation, using
the \cite{Caldwell1981} rotation curve for constraining the dark halo density, and deriving the eccentricities of the Einasto profiles for each thin disc sub-component from the collisionless Boltzmann equation, assuming dynamical statistical equilibrium. The resulting values for the Einasto eccentricities are given in Tables  \ref{density} and \ref{eccentricities}. 

\begin{table*}
\caption{Thin disc local densities $M_\odot / pc^3$. Contribution to the total dynamical mass of the 7 sub-populations (see Sect. 2.1) for each one of the model variants in Figure 3.      }  
\label{density}       
\centering          
\begin{tabular}{c c c c c l l l l }     % 9 columns 
\hline\hline       
                      % To combine 4 columns into a single one 
Sub-population & Age (Gyr)& DM & HRV & DBV & DCV & HRVB & SV& DAV   \\ 
\hline                    
   1  & 0-0.10 & 0.00131 & 0.00117 & 0.00147 & 0.00207 & 0.00133 & 0.00113 & 0.00128 \\  
   2  & 0.10 -1 &0.00541   & 0.00504 & 0.00639& 0.00808 & 0.00589 &  0.00468& 0.00530 \\
   3  & 1-2 & 0.00427    & 0.00404 & 0.00499& 0.00557 & 0.00483 & 0.00382 & 0.00417  \\
   4 & 2-3 & 0.00291   & 0.00284& 0.00351& 0.00339& 0.00341 & 0.00275& 0.00292 \\
   5  & 3-5 &0.00496  & 0.00502 & 0.00580& 0.00485 &0.00596 & 0.00496 & 0.00498 \\
   6  &5-7& 0.00510   & 0.00526 & 0.00601& 0.00393 & 0.00620& 0.00535 & 0.00506 \\
   7  & 7-10 &0.00944 & 0.00997 &  0.01122& 0.00543  & 0.01183 & 0.0103 & 0.00953 \\
   Total thin disc  &  &0.0334 & 0.0333& 0.0394& 0.0333 & 0.0395 &0.0330 &0.0332 \\
\hline                  
\end{tabular}
\end{table*}

\begin{table*}
\caption{Thin disc eccentricities of the 7 sub-populations (see Sect. 2.1) for each one of the model variants in Figure 3.}             
\label{eccentricities}      
\centering          
\begin{tabular}{c c c c c l l l l }     % 9 columns 
\hline\hline       
                      % To combine 4 columns into a single one 
Sub-population & Age (Gyr)  & DM & HRV & DBV & DCV & HRVB & SV & DAV  \\ 
\hline                    
   1  & 0-0.10 & 0.0140 & 0.0140 & 0.0140 & 0.0140 & 0.0140 & 0.0140 & 0.0140 \\  
   2  & 0.10-1 & 0.0205   & 0.0204 & 0.0197& 0.0210 & 0.0196 &  0.0205 & 0.0231 \\
   3  &1-2& 0.0292    & 0.0292 & 0.0281& 0.0299 & 0.0280 & 0.0292 & 0.0327  \\
   4  &2-3&0.0441   & 0.0440& 0.0426& 0.0450 & 0.0424 & 0.0441& 0.0489  \\
   5  &3-5& 0.0565   & 0.0564 & 0.0547& 0.0576 &0.0545 & 0.0565 & 0.0624 \\
   6  &5-7&0.0642   & 0.0641 & 0.0622& 0.0654 & 0.0619& 0.0642 & 0.0707 \\
   7  &7-10 &0.0647 & 0.0645 &  0.0627& 0.0659  & 0.0624 & 0.0646 & 0.0712 \\
\hline                  
\end{tabular}
\end{table*}

\subsection{Initial mass function}\label{IMFs}

\begin{figure}
   \centering
 \includegraphics[width=\hsize]{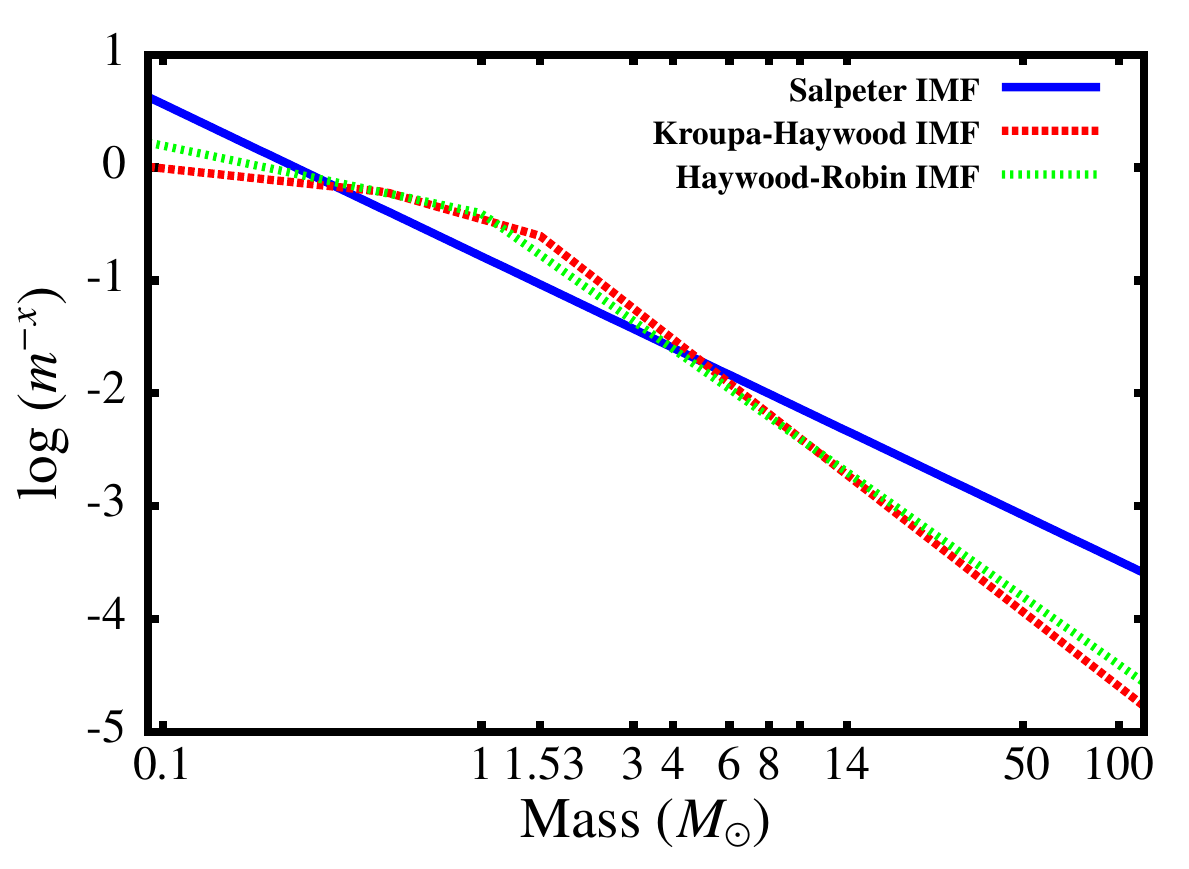}
      \caption{Three tested IMFs in the range between 0.09 and 120 $M_\odot$ normalized. The blue solid line is for Salpeter IMF, the red thick dashed line is for Kroupa-Haywood, and the green thin dashed line is for Haywood-Robin. We note  how for a fix total amount of mass, Salpeter IMF is the IMF that generates less stars in the interval $\approx 0.5$ to $\approx 6 M_\odot$, but more stars at $M> 6 M_\odot$ and at $M<0.5 M_\odot $}
         \label{IMF}
\end{figure}

 In Figure \ref{IMF}, we present the normalised IMFs that we proposed to  test. They are described using the classical analytical approximation $\xi(m)$:

\begin{equation}
\frac{dN}{dm}=\xi(m)=k \cdot m^{-\alpha} = k \cdot m^{-(1+x)}
\end{equation}

where N is the number of stars, m is the mass (in $M_\odot$), and k is the normalisation constant. We propose to check the two IMFs that better fit the Tycho-2 data \citep{Czekaj2014}. The slope of these IMFs at intermediate masses are on the upper limit values found in the literature (see \citealt{KroupaSci2002}). Additionally, to cover most of the range of variation of the slope, we include the classical IMF of \cite{Salpeter1955} with $\alpha = 2.35,$ as representative of the lowest values. In Table 3, we present the slopes and limiting masses for the tested IMFs.

In Figure \ref{IMF}, we present the three tested IMFs normalized in the range between 0.09 and 120 $M_\odot$. The normalization has been done in terms of mass. To get the mass locked within each mass interval, one must solve $\int{m \cdot \xi(m) dm}$ for each mass range. The sum of the obtained result for each mass interval is then normalized to one using the continuity coefficients $K_i$ as $\sum_i K_i \cdot \int^{m_{i+1}}_{m_i} m \cdot \xi(m) dm= 1$, i being the index of the mass interval considered (i=1, 3), see \cite{Czekaj2014} Section 2.2.1. Looking at Figure \ref{IMF}, we can anticipate the general lines of the simulations. Salpeter IMF will produce more stars in the range between 0.09 and 0.5 $M_\odot$ while in the range between 0.5 to about $5 M_\odot$, it will produce less stars than the two other tested IMFs. From about $5 M_\odot$ to $120 M_\odot,$ Salpeter IMF will be the IMF that  produces more stars. If we take a look at the tested IMFs in the Cepheid mass range, it is clear that Salpeter will dedicate more mass to the Cepheid production than the other two IMFs. We note, however, that the Salpeter IMF generates less low-mass Cepheids (in the range $\approx 3M_\odot$ to $\approx 6M_\odot$) and much more high-mass Cepheids ($M>6M_\odot$) than the other tested IMFs. Kroupa-Haywood IMF will produce a few more Cepheids than Haywood-Robin IMF.

\begin{table*}
\caption{Slopes and mass limits for the tested IMFs. The $M_1$, $M_2$, $M_3,$ and $M_4$ are the limiting masses (when necessary) and the $\alpha_1$, $\alpha_2$ and $\alpha_3$ are the corresponding slopes. The values of the $M_1$ and $M_4$ are fixed according to the limiting masses of the evolutionary tracks.}             % title of Table
\label{IMFtable}      % is used to refer this table in the text
\centering                          % used for centering table
\begin{tabular}{c c c c c c c c}        % centered columns (4 columns)
\hline\hline                 % inserts double horizontal lines
IMF & $M_1$ & $\alpha_1$ & $M_2$ & $\alpha_2$ & $M_3$ & $\alpha_3$ & $M_4$ \\    % table heading 
\hline                        % inserts single horizontal line
   Salpeter & 0.09 & 2.35 & - & 2.35 & - & 2.35 & 120 \\            Haywood-Robin & 0.09 & 1.6    & 1.0 & 3.0 & - & 3.0   &  120\\
   Kroupa-Haywood & 0.09 & 1.3 & 0.5    & 1.8 & 1.53 & 3.2 & 120 \\
   
\hline                                   %inserts single line
\end{tabular}
\end{table*}

\section{Classical Cepheid observational data}\label{obs}

Our strategy requires us to compare well-defined samples which are complete up to a limit in apparent magnitude. Whereas this is trivial for simulated BGM samples, observational data have to be treated rigorously. Currently, the most complete Cepheid catalogues with good Cepheid variability classification are the Berdnikov catalogue \citep{Berdnikov2008} with 577 stars and the DDO catalogue \citep{Fernie1995} with 509 stars, with both catalogues being compilations of photometric data for known Cepheids. We  tested that 95\% of Cepheids up to V=9 are contained in both catalogues, thus in the present work we have used the photometric data from the Berdnikov catalogue (for Cepheids up to V=9) since it is the most up-to-date one. For fainter magnitudes, we  used the  ASAS Catalogue of Variable Stars (hereafter ACVS) from \cite{Pojmanski2002} and \cite{Pojmanski2006}. The telescopes for this survey are installed in the southern hemisphere from where stars with declination $\delta  \leq 29$ can be observed. ACVS contains 809 stars classified exclusively as classical Cepheids. Whereas the quality of the light curves of the variable stars in the Bernikov catalogue ensures a good classification of the classical Cepheids, the light curves of the variable stars in ACVS are less accurate. Thus the Classical Cepheids sample of ACVS could contain some contaminants from other variable types. To minimize  the contamination, we work only with those Cepheids in ACVS that are concentrated in the Galactic plane.

The type II Cepheids, the old, low-mass counterpart to the classical Cepheids are believed to belong to the thick disc.  They are difficult to disentangle from classical Cepheids. For the brightest Cepheids, for which large amounts of photometric data is available and the chemical composition is known (e.g. \citealt{2002A&A...392..491A,2007A&A...467..283L}), the amount of contamination from type II Cepheids is negligible. For the faintest Cepheids of our sample, since we  concentrate on the Galactic plane, where the contribution of the thick disc is small, the contamination from type II Cepheids is not significant for our study.

Classical Cepheids are variable stars with  large amplitude variations in visual magnitude. As a result,  caution is necessary when defining the photometric parameters setting the limiting magnitude. The Berdnikov catalogue provides both the visual magnitude at maximum ($V_{max}$) and minimum brightness ($V_{min}$). ACVS provides the  visual magnitude at maximum brightness ($V_{max}$) and the amplitude ($\delta V$). To accurately determine the mean magnitude of a Cepheid,  template lightcurves should be used. But, as we are conservative when selecting the limiting magnitude for the completeness of the catalogues, we can approximate  $V_{mean}$ by $(V_{max}+V_{min})/2$  for Berdnikov data and by ${V_{max}+\delta V/2}$ for ACVS. With this definition in mind, several considerations arise when comparing simulated and observed data.

Since our BGM simulations do not include brightness variability information (see Section 2.1 and 2.2.), it is appropriate to consider the magnitude from the evolutionary tracks as the one associated to the mean intrinsic brightness of the star. The light curve of the Cepheids can be assumed as being symmetric at first approximation; then the probability of finding a Cepheid in any point of its period between $V_{max}$ and $V_{min}$ is uniform.

Owing to Cepheid variability, one might wonder whether a bias in the star counts similar to Malmquist bias  could be introduced when cutting the sample in $V_{mean}$ apparent magnitude. 
To quantify this effect, we have taken the full Berdnikov data, and assigned a random phase to each star in the catalogue. This process was done by assigning a V magnitude to each star in the range $[V_{min}$,$V_{max}]$ with a uniform probability. This process was repeated to generate 10 000  realisations of the catalogue and a cut to $V=9$ was applied in each realisation. These 10 000 realisations gave us a mean number of counts of $141 \pm 2$  Cepheids up to $V=9$. Then we verified that the same number of Cepheids (141 stars) was obtained when a cut at $V_{mean}$ = 9 was applied. From this test, we prove that the comparison of observed and simulated data can be done using $V_{mean}$ as the limiting magnitude.  

The next step was to set up the faint-end apparent magnitude completeness limit values for each catalogue. For both Berdnikov and ACVS catalogues, this limit  was evaluated as in \cite{Monguio2013}. The limiting magnitude was computed as the mean of the magnitudes at the maximum peak star counts in a magnitude histogram, and its two adjacent bins, before and after the peak, weighted by the number of stars in each bin. In \cite{Monguio2013}, it was estimated that the limiting magnitude computed with this method provides the 90\% completeness limit. They confirmed it by using complete catalogues. Following this strategy, we obtain the 90\% completeness limit of Berdnikov catalogue at $V=9.5$, for ACVS we obtained the  90\% completeness at $V=12.4$. From these results it is reasonable to consider the Berdnikov catalogue complete at $V=9$ and the ACVS catalogue complete at $V=12$. To summarize, our observed sample has 141 classical Cepheids from full sky with visual magnitude up to $V=9$ (Bernikov catalogue) and 279 Cepheids in the magnitude range $9 \leq V \leq 12$  with $\delta  \leq 29$ and $|b|\leq 10$ (from ACVS). This observational constraint can be well modelled in our simulated BGM sample.

\section{Statistical tools for IMF's evaluation}\label{methods}

Three different statistical methods are used to search for the best IMF fitting the observations.  As will be seen, the information provided by each of them is fully complementary. Furthermore, whereas a unique method could converge to the non-optimal solution, a robust conclusion is obtained  when consistency among the three is obtained. 

\subsection{Absolute Cepheid counts}

Our first evaluation method is as simple as comparing the total number of simulated Cepheids versus the observations. This method allows us to test which IMF and model variant is able to reproduce the total number of classical Cepheids up to a given limit in apparent magnitude. This strategy shows us how all the machinery of the BGM, which incorporates most of the current knowledge about the Milky Way, is able to generate the observed amount of a certain type of stars in a specific evolutionary stage. It is the first time that the BGM is used to test such a specific population as  classical Cepheids. 

Each one of the model variants in Figure \ref{variants} was simulated ten times,  and the mean of the resulting star counts was computed. To quantify the differences between model and data, we use a basic estimator, $\chi^2 = {\left(N_{obs}-N_{simu}\right)^2}/{N_{obs}}$. This exercise is done for each one of the two extinction models considered, resulting in a total of  280 simulations on the MareNostrum supercomputer. 
This evaluation method has the following drawback: it could be possible to find a combination of parameters that works properly when fitting the Cepheid observational data, but it could fail when trying to fit the observational data of other stellar populations or the whole sky. A given IMF could be able to reproduce the absolute number of Cepheids up to a given limiting magnitude, but this does not prove its goodness in a general sense. To overcome this drawback,   in Section 4.2 we introduce the reduced likelihood test to be applied to a full sky sample (in this case Tycho-2 data) and, in Section 4.3, a probabilistic approach that involves  all populations as well as Cepheid data.

\subsection{Reduced likelihood applied to Tycho-2 data}

 This method was designed to search for the IMF and model variant that gives a better fit with Tycho-2 data in the region of low latitudes $|b| \leq 10^{\circ}$ . As mentioned, we are searching for the IMF that best fits the Cepheids but also the full thin disc population. The Galactic plane was selected because, as known, the youngest population, and thus  classical Cepheids,  are well concentrated  in this region. The reduced likelihood for Poisson statistics has been selected to undertake this work as  described in \cite{Bienayme1987}. The absolute value of this likelihood needs to be understood as a good distance estimator to evaluate the differences between the simulations and the observations in terms of star counts. As pointed out in \cite{Bienayme1987}, this method avoids the bias introduced by the chi-square fit, at least for small numbers. 

For each model variant (Figure \ref{variants}), a distance is computed between simulated and observed star counts, taking the absolute value of the following expression:

\begin{equation}
L_r = \sum_{i=1}^N q_i \cdot (1-R_i+ln(R_i))
,\end{equation}

where $L_r$ is the reduced likelihood for a Poisson statistics (\citealt{Kendall1973,Bienayme1987}) and $q_i$ and $f_i$ the number of stars in the data and the model respectively. $R_i$ is defined as the quotient between both ($R_i = f_i / q_i$). This reduced likelihood becomes zero when the simulation and the observations have the same number of stars in each bin, and $L_r=0$ is its maximum value. $|L_r|$ can be understood as a metric for the distance between simulations and observations in terms of star counts in a 2D grid. The smaller the value of  $|L_r|$, the closer the simulated data is to the observational data.

This reduced likelihood is applied to the 1D colour $(B-V)_T$ distribution (1D plots like the ones used in \citealt{Czekaj2014}) and to the 2D case of the colour-magnitude diagram distribution (hereafter CMD) used in \cite{Robin2014}.

\subsection{The probabilistic Bayesian approach}\label{Bayes}

This third evaluation method has been developed to simultaneously use, in a Bayesian probabilistic approach, data from both Cepheids and all stellar populations found in the disc. It is also  applicable to the full sky star count distribution.

We want to quantify how good a given IMF is able to reproduce the probability to find a Cepheid each time a star is observed. As known this probability depends on the apparent limiting magnitude of the sample.  Hereafter, for simplicity, this probability is called the Cepheid fraction.
We aim at quantifying, within the tolerance interval, the probability that our model variant, which is being evaluated, has the same Cepheid fraction as the observations. This strategy is equivalent to the one recently proposed by \cite{Downes2015}. These authors applied the method to establish the number fraction of stars with circumstellar discs among low mass stars and brown dwarfs. The method imposes  choosing a tolerance threshold when comparing simulated and observed Cepheid fraction (e.g. 15\%). This threshold defines a tolerance interval in the 2D probability space established by these two Cepheids fractions.

Our problem is a two-state problem: for a given observed star, either it is a Cepheid or it is not, i.e. we have a so-called success if the observed star is a Cepheid and a so-called failure if it is not. As already known, this can be described by the binomial distribution. 

Let $f^{Obs}_{Cep}$ and $f^{sim-IMF}_{Cep}$ be the observed and the simulated Cepheid fraction. Since these probabilities are independent, we can write the full posterior probability distribution function in the 2D space as
 
\begin{equation}
\label{eq2}
P \left ( f^{obs}_{Cep}, f^{sim-IMF}_{Cep} | data \right ) = P \left ( f^{obs}_{Cep} | data \right ) \ast P \left (f^{sim-IMF}_{Cep} | data \right )
,\end{equation}

where $P \left ( f^{obs}_{Cep} | data \right )$ and $P \left (f^{sim-IMF}_{Cep} | data \right )$ are binomial distributions, thus each of them can be computed following 

\begin{equation}
\label{eq3}
P \left ( f_{Cep} | data \right )=  \left ({(f_{Cep})}^{N_{Cep}} \cdot (1-f_{Cep})^{N_{tot}-N_{Cep}} \right )
.\end{equation}

Substituting these expressions in Equation \ref{eq2} and adopting a uniform prior, the posterior full 2D probability can be expressed as

\begin{displaymath}
P \left ( f^{obs}_{Cep}, f^{sim-IMF}_{Cep} | data \right )= C \cdot \left ({(f^{obs}_{Cep})}^{N^{Obs}_{Cep}} \cdot (1-f^{obs}_{Cep})^{N^{Obs}_{tot}-N^{Obs}_{Cep}} \right ) \ast 
\end{displaymath}
\begin{equation}\label{eq4}
 \ast \left({(f^{sim-IMF}_{Cep})}^{N^{sim-IMF}_{Cep}} \cdot (1-f^{sim-IMF}_{Cep})^{N^{sim-IMF}_{tot}-N^{sim-IMF}_{Cep}} \right )
,\end{equation}

where C is the normalisation constant. Following \cite{Downes2015}, Equation 3, the integral, over the full tolerance area, of the posterior probability distribution function (Equation \ref{eq4}) gives the probability that the observations and the model variant have the same Cepheid fraction.

As mentioned, this method can be applied to full sky data. In our case, owing to Cepheid observational constraints (see Section 3), it will be applied to the full sky data for samples up to V=9 and to  $\delta  \leq 29^{\circ}$ and $|b|\leq 10^{\circ}$ area for the magnitude range $9 \leq V \leq 11$. This upper limit of V=11 is imposed, in this case, by the completeness of the Tycho-2 catalogue \citep{Czekaj2014}.

\section{Results}\label{results}

The three evaluation methods described in the previous section have been applied to Cepheids and Tycho-2 data. We note that each of these methods is dominated by a specific range of masses. Whereas in the absolute Cepheid count method the dominant masses are in the range 3-15 $M_\odot$, that is in the mass range of our Cepheids, in the likelihood method the most dominant objects are the low mass stars in Tycho-2 catalogue. And, in a complementary way, the resulting probabilities in the Bayesian  method are influenced by both low and intermediate mass.

\begin{figure}
   \centering
   \includegraphics[width=\hsize]{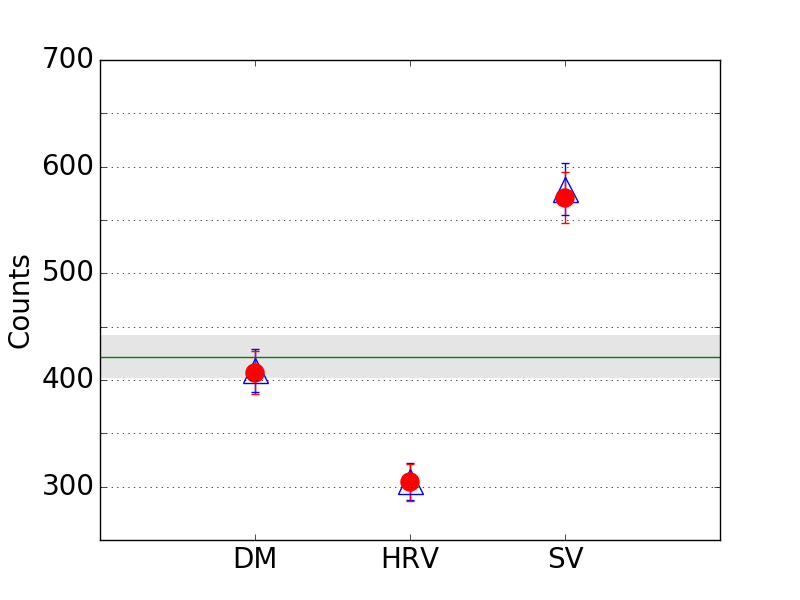}
      \caption{Testing the IMF. Cepheid counts for the complete region up to $V=12$. Up to $V=9$ the observational Cepheid catalogues are considered as being complete for the whole sky while for the interval $9 < V \leq 12$ they are supposed to be complete for $\delta \leq 29$. In the interval $9 < V \leq 12,$ an additional cut ($|b| \leq 10$) is applied to avoid contamination of the observational sample. The green line indicates the observational counts, the grey region is the region within $1 \sigma$. Filled red dots are for the simulations with the \cite{Marshall2006} extinction model, while blue triangles are simulations with the \cite{Drimmel2001} extinction model. Error bars are due to Poisson noise.  We note  how the DM, that is Kroupa-Haywood IMF, is the variant that better fits the observational data, while HRV, that is Haywood-Robin IMF and SV (Salpeter IMF), are more than $5 \sigma$ away from the observational data.}
         \label{CphCounts}
   \end{figure}
   
   \begin{figure}
   \centering
   \includegraphics[width=\hsize]{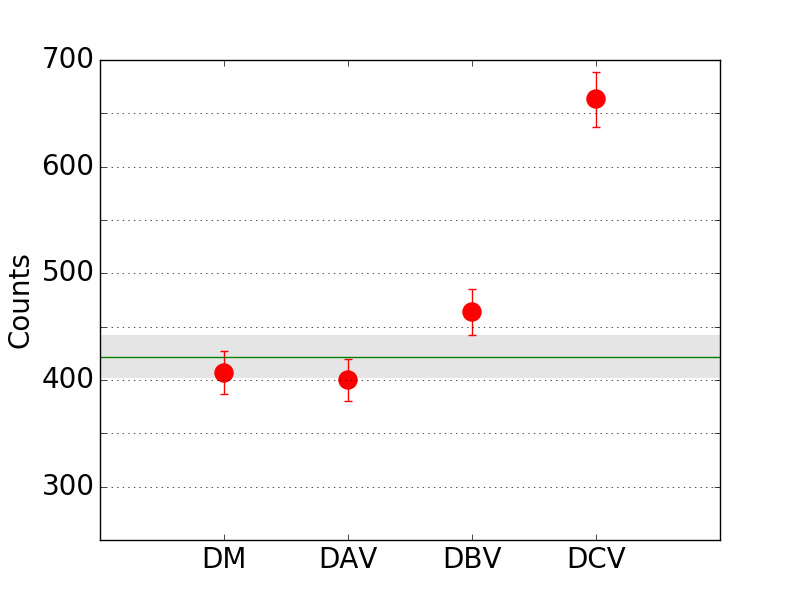}
      \caption{Testing variations on radial scale length, stellar density at Sun position, and SFH. Cepheid counts for the same completeness regions as Figure \ref{CphCounts}. The green line indicates the observational counts, the grey region is the region within $1 \sigma$. Notice how DAV (changed scale length) and DM (default model) are really close to the observational data. The simulations have been made taking into consideration the \cite{Marshall2006} extinction model. Hence reasonable changes in the radial scale length have small effects in the total Cepheid counts. DBV (local stellar density variant) is still close to observational data. As expected, a change in SFH (DCV) is critical for the comparison of Cepheid counts in absolute terms.  }
         \label{testCphcounts}
   \end{figure}

In Figure \ref{CphCounts}, we present the comparison of the absolute Cepheid counts as part of the first evaluation method. We can see how the DM, which uses Kroupa-Haywood IMF, nicely fits the observational data with a $\chi^2=0.4,$ while Haywood-Robin IMF (HRV) and Salpeter IMF (SV) are more than $5\sigma$ away from the observed data with $\chi^2=32$ and $\chi^2=55,$ respectively. Thus, our first evaluation method places the Kroupa-Haywood IMF as the best to reproduce the absolute Galactic Cepheid counts. We emphasize that, although \cite{Czekaj2014} showed that the extinction model can play a significant role in star count comparisons, our analysis shown in Figure 5 does not critically depend on it. However it should be noted that Marshall's extinction model covers only about half of the Galactic plane ($|l|\leq 100^{\circ}$), so any difference between extinction models should come from this area. In a similar way, we have checked that, in terms of star count computation, the already reported Marshall underestimation of extinction (e.g. \citealt{Czekaj2014}) at short heliocentric distances compensates the underestimation of Drimmel's model with regards to Marshall's model at distances that are larger than about 1 kpc. We would need more accurate extinction maps (for example from future Gaia Data) to treat the absolute Cepheid counts in the solar neighbourhood more robustly.

As mentioned in Section 2.5, we want to evaluate and quantify the impact in previous results when changing critical ingredients, such as the radial scale length of the thin disc, the stellar density at Sun position, and the SFH. In Figure 6, we present a comparison between observational data and model variants for which these parameters have been changed. To quantify the impact of a change in the radial scale length, we need to look at the differences between the DAV model, with a radial scale length of 2170pc and the DM, where a radial scale length of 2530pc is used. We note in Figure 6 that this difference is small and both DM and DAV fit  the observational data properly. Since the stellar density at the position of the Sun is fixed in BGM, a change in the radial scale length means, for simulations in the Solar neighbourhood, a change in the stellar density distribution towards the Galactic centre, compensated for by an opposite change towards the Galactic anticentre.

As expected, an increase in the stellar density at Sun position from $0.033 M_\odot pc^{-3}$ in the DM to $0.039M_\odot pc^{-3}$ in the DBV Model variant produces an increase of the Cepheid counts, however this deviation is only at 1-2 sigma from the observed values. Finally, we  quantified the effects of considering a constant SFH instead of the decreasing SFH proposed by \cite{Aumer&Binney2009} and used in the DM. We  verified that the impact of considering a constant  SFH is critical and simulations deviate from the observed star counts by more than $5\sigma$. Although not shown in the figure, model variant HRVB with Haywood-Robin IMF and local stellar density of 0.039 is still generating too few Cepheids (352) at more than $3\sigma$ from observational data with $\chi^2=11.6$,  indicating that the IMF effect is dominant over the local stellar density. To conclude, the absolute count method applied here demonstrates that, even with reasonable changes in stellar density and radial scale length, the Kroupa-Haywood IMF is still the best option to reproduce observed Cepheid counts.

In Tables \ref{BV} and \ref{CMD}, we present the results when applying the reduced likelihood method (see section 4.2) to the 1D $(B-V)_T$ colour distribution and 2D Colour-Magnitude Diagram respectively. In this analysis, we test the model variants presented here using Tycho-2 data (all populations) in the Galactic plane. To compare our results with \cite{Czekaj2014}, we have also computed the reduced likelihood for Model A and B of the mentioned paper. The results (see Tables 4 and 5) confirms that Model A, using Haywood-Robin IMF, fits  the Galactic Plane regions better than Model B as reported in \cite{Czekaj2014}. We have obtained smaller values of $|L_r|$ for our DM and DAV model variant than the values obtained for Model A and Model B, which means that our best models are improving the results of Model A and B of \cite{Czekaj2014} when fitting BGM with Tycho-2 data in the Galactic plane. This improvement is probably due to both the new age range assigned to the sub-populations 1 and 2 of the thin disc (Section 2.4) and the new strategy adopted to apply photometric transformation \footnote{Johnson V magnitudes have been transformed to Tycho-2 ($V_T$) following the strategy proposed in \cite{1997ESA} (see Vol. 1, Table 1.3.4). Transformations have been applied before adding the extinction and, for unresolved binary systems, before merging fluxes.}. For simplicity, we do not list the $|L_r|$ values corresponding to all the model variants presented in Figure 3 in these tables, only the best of our models are shown. We note that, although in the 1D case the HRVB model variant, with the Haywood-Robin IMF, has the second smallest $|L_r|$, this is no longer the case when we consider the 2D CMD, where the best models  use Kroupa-Haywood IMF.

This evaluation method leads us to favour Kroupa-Haywood as the best IMF for the Galactic plane. Furthermore, we can  confirm that this model variants improve the fit with the observational data with respect to the ones on \cite{Czekaj2014}.  

\begin{table}
\caption{Absolute values of the reduced likelihood for the models fitted to Tycho-2 data in the Galactic plane using colour distributions.  \textbf{Notes}. $|L_r|$ is a good distance estimator between the simulations and the observations in terms of star counts in a 2D grid to quantify its differences. Smaller values corresponds to better fits.}             % title of Table
\label{BV}      % is used to refer this table in the text
\centering                          % used for centering table
\begin{tabular}{c c c}        % centered columns (4 columns)
\hline\hline                 % inserts double horizontal lines
Thin Disc Model &  Extinction Model & |$L_r$| \\    % table heading 
\hline                        % inserts single horizontal line
   Model A \cite{Czekaj2014} & Marshall & 6350  \\      % inserting body of the table
   Model B \cite{Czekaj2014} & Marshall & 15653  \\
   HRVB & Marshall & 5130 \\
   Default Model (DM) & Marshall &5302  \\
   DAV & Drimmel & 4300 \\
  
\hline                                   %inserts single line
\end{tabular}
\end{table}

\begin{table}
\caption{Absolute values of the reduced likelihood for the models fitted to Tycho-2 data in the Galactic plane using colour-magnitude diagrams.  \textbf{Notes}. $|L_r|$ is a good distance estimator between the simulations and the observations in terms of star counts in a 2D grid to quantify its differences. Smaller values corresponds to better fits.}             % title of Table
\label{CMD}      % is used to refer this table in the text
\centering                          % used for centering table
\begin{tabular}{c c c}        % centered columns (4 columns)
\hline\hline                 % inserts double horizontal lines
Thin Disc Model & Extinction Model & |$L_r$|  \\    % table heading 
\hline                        % inserts single horizontal line
   Model A \cite{Czekaj2014} & Marshall &9037  \\      % inserting body of the table
   Model B \cite{Czekaj2014} & Marshall &18357  \\
   HRVB & Marshall & 7708 \\
   Default Model (DM)& Marshall &6936  \\
   DAV & Drimmel &5645 \\

\hline                                   %inserts single line
\end{tabular}
\end{table}

    \begin{figure}
   \centering
   \includegraphics[width=\hsize]{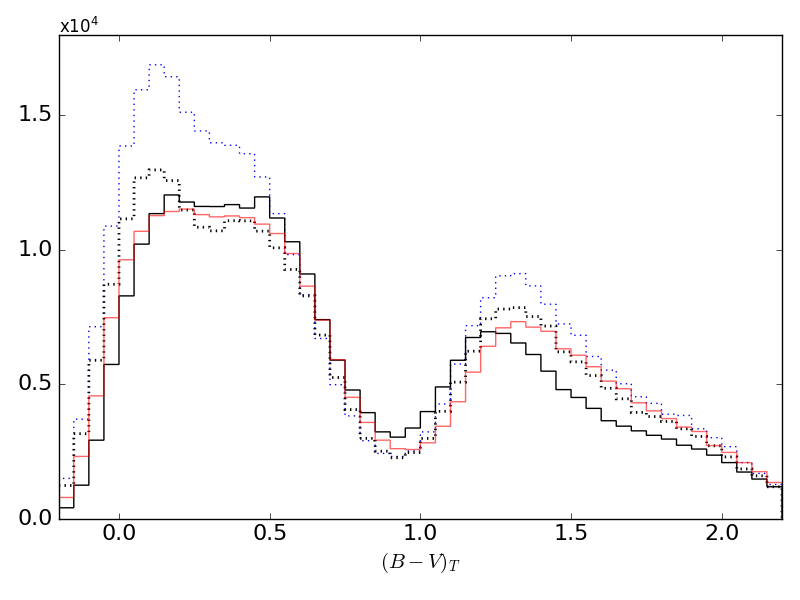}
      \caption{Colour $(B-V)_T$ distribution for the Galactic plane $(|b| \leq 10)$. The black solid thick line is for Tycho-2 catalogue, the dotted blue and black lines are respectively for models A and B of \cite{Czekaj2014}, the red solid thin line is for our model variation DAV, which gives the best fit with the observational data.}
      \label{BVTGP}
   \end{figure}
   
        \begin{figure}
   \centering
   \includegraphics[width=\hsize]{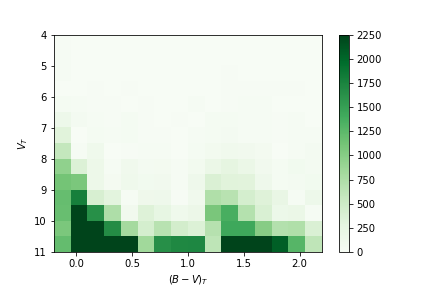}
      \caption{Absolute differences in star counts in the colour-magnitude diagram between Tycho-2 data and Model B from \cite{Czekaj2014} in the Galactic plane.}
      \label{ResB}
   \end{figure}
   
        \begin{figure}
   \centering
   \includegraphics[width=\hsize]{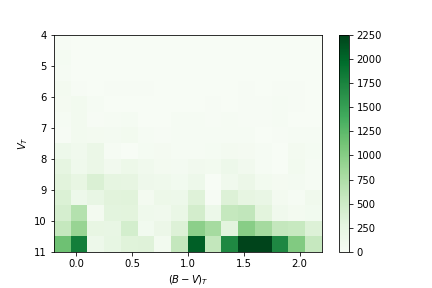}
      \caption{Absolute differences in star counts in the colour-magnitude diagram between Tycho-2 data and DAV variant simulation in the Galactic plane.}
      \label{ResDAV}
   \end{figure}

In Figure \ref{BVTGP}, we present the colour $(B-V)_T$ distribution in the Galactic plane $(|b|\leq 10 ^{\circ})$ for Tycho-2 (solid-black) data and for simulations using: (1) Model A from \cite{Czekaj2014} (dotted-black), (2) Model B from \cite{Czekaj2014} (dotted-blue) and (3) Our DAV model variant (solid-red), our best fit model variant. As can be seen, the blue peak around $(B-V)_T \approx 0.15$ that does not match Tycho-2 data with old Model A and B \citep{Czekaj2014} is now perfectly well reproduced when our new DAV model variant is considered.

In the red peak of the colour distribution, it can be seen that all models are shifted by about 0.05 magnitudes from the observed data. This is a long-standing problem, most probably related to the stellar atmosphere models used in the simulation or to the photometric transformation for red giants. As this does not impact our present study, we will consider it in a future paper.

In Figures \ref{ResB} and \ref{ResDAV}, we present the absolute residuals in star counts between model and Tycho-2 data in the Galactic plane in the colour-magnitude distribution. Figure \ref{ResB} is created using Model B from \cite{Czekaj2014} whereas, in Figure \ref{ResDAV}, our DAV variant is used. We note how this model variant improves the results in the overall diagram and even more in the blue region. However, as commented above, some significant differences  still remain, specifically in the faint red region.

\begin{figure*}
   \centering
   \includegraphics[width=\hsize]{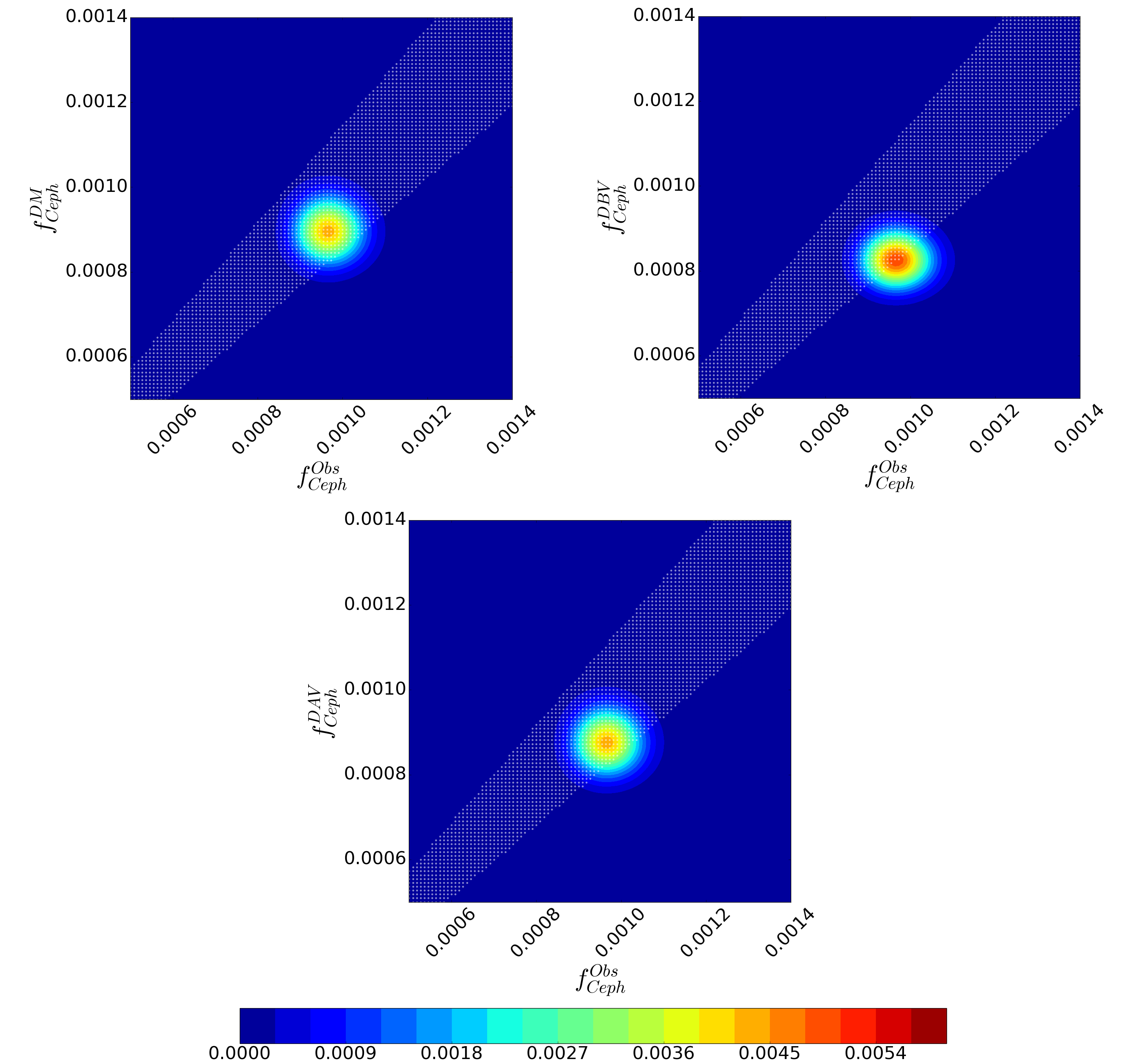}
      \caption{Full 2D posterior probability distribution function. The white stripe shows the tolerance interval region. The plot is for the three cases in Figure \ref{testCphcounts} that needs a disambiguation: Top left panel: DM; Top right panel: DBV (different local density); Bottom panel: DAV (best fit model with new scale length). Note here how DM and DAV are almost completely inside the tolerance region, while DBV is half out. DM has an $ \approx 85\%$ of probability to have the same observed Cepheid fraction as the Milky way up to magnitude $V=11,$ while DBV is just in the $\approx 50\%$ } 
         \label{Bayes1}
   \end{figure*}

      \begin{figure}
   \centering
   \includegraphics[width=\hsize]{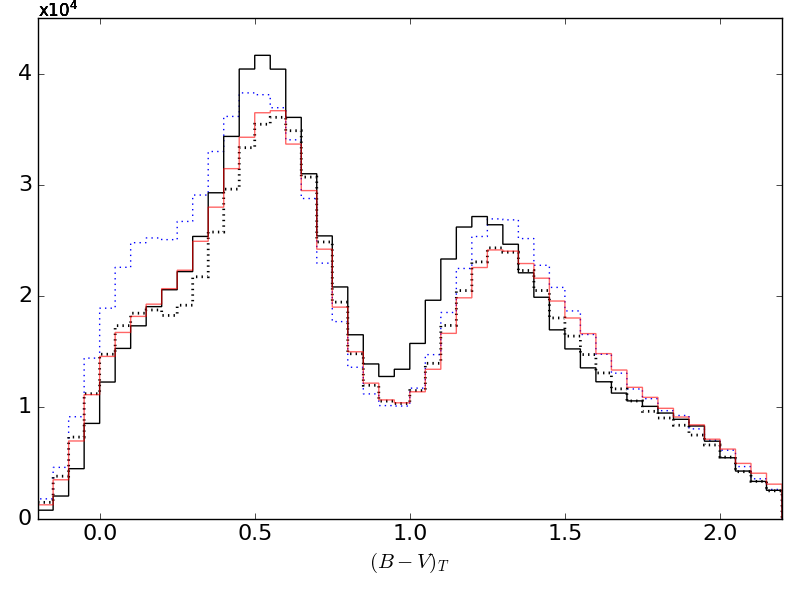}
      \caption{Colour $(B-V)_T$ distribution for the whole sky. The black solid thick line is for Tycho-2 catalogue, the dotted blue and black lines are, respectively, for models A and B of \cite{Czekaj2014}, the red solid thin line is for our model variation DAV.}
      \label{BVTallsky}
   \end{figure}

As a final step, and to add statistical robustness to previous conclusions, we  applied the probabilistic Bayesian approach (Section 4.3), to study the Cepheid fraction, which simultaneously combined  Cepheids and Tycho-2 data.
In Figure \ref{Bayes1} we show, for those model variants differing by less than 1-2 $\sigma$ from observational data in Figures \ref{CphCounts} and \ref{testCphcounts}, the full 2D posterior probability distribution function of the Bayesian problem. All  models plotted here are generated using Kroupa-Haywood IMF. The white region is the tolerance region, the full posterior 2D distribution function is integrated over the tolerance region searching for the IMF giving higher probability to reproduce the observed Cepheid fraction up to $V=11$. Whereas the DM has a $\approx 85\%$ probability of having the same observed Cepheid fraction as the Milky Way, up to V=11, the DAV has a $\approx 78\%$ while DBV is just in the $\approx 50\%$. The HRVB, with Haywood-Robin IMF gives us only a probability of $\approx 60\%$ (not plotted here). We checked that all the other model variants using Salpeter or Haywood-Robin IMF always give  probabilities smaller than $\approx 30\%$. The Kroupa-Haywood IMF is the tested IMF with the highest probability to reproduce the observed Cepheid fraction. To sum up,  the Kroupa-Haywood IMF gives the best results out of all the  evaluation methods used.

For completeness, in Figure 11 we compare the $(B-V)_T$ colour distribution for the whole sky Tycho-2 data with our best fit Model variant (DAV) implemented here. We note how the fit of our model with the whole sky young population in the blue peak of the (B-V) histogram has significantly improved. %Again, as can be seen, the red peak is still slightly shifted by about 0.05 magnitudes as in \cite{Czekaj2014}, as commented above this shift could be associated  to the stellar atmosphere model used or the photometric transformation for Red Giants.

\section{The IMF in the solar neighbourhood}\label{discussion} 

All methods and data evaluated in the previous section point towards the Kroupa-Haywood IMF as the one that best fits Cepheids and Tycho-2 data. This IMF is described with a simplified analytical form with three truncated power laws. Using field stars and Cepheids, our fits point towards a slope of about $\alpha = 3.2$ at intermediate masses, and excludes the flatter values of $\alpha = 2.35$ of Salpeter IMF. We now want to discuss  the scope of these results both in terms of star formation environment and time evolution in the Galactic disc. Our results were  obtained using  Galactic Cepheids  at all Galactic longitudes located  up to $\approx 2 $kpc from the Sun. Thus, the derived IMF reflects the mass distribution of the formation episodes that took place in the last 200 Myrs (upper limit of the Cepheid age) over this region. Our IMF should be understood as a composite IMF, as described in  \cite{Kroupa2013}, one of the best references and review  in this field recently published. Instead of being an IMF derived from an individual cluster or association, our IMF is the sum of all stellar IMFs from the star formation episodes in the local thin disc environment.

The comparison of our results ($\alpha = 3.2$) with those  in the literature is complex. To begin with, studies using clusters and OB associations (e.g. \citealt{Massey1998}) show that, for stars more massive than the Sun, the IMF can be well approximated by a single power-law function with the Salpeter index $\alpha=2.35$. As mentioned in \cite{Kroupa2003}, there is a discrepancy between the slope of the IMF obtained using field stars ($\alpha_{field}$) and the slope of the IMF obtained from stars belonging to a cluster ($\alpha_{cluster}$). This discrepancy $(\alpha_{field} > \alpha_{cluster})$ can be explained by the fact that low mass clusters are more abundant than the most massive clusters, then the contribution of the low mass clusters to the field stars is higher. The abundant low mass clusters do not have massive stars, while the rare massive clusters do, and this leads to a steepening of the composite IMF $(\alpha_{field} > \alpha_{cluster}),$ which is a sum of all the IMFs in all the clusters that spawn the Galactic field population \citep{Kroupa2003}.

Several other studies have derived the IMF using field stars. We should mention the classical work of \cite{Scalo1986} who derived a slope of $\alpha=2.7$ for $M> 1M_\odot$. To compare studies with our results one has to keep in mind that the complexity grows owing to the different ingredients assumed in each case. Critical parameters such as  the SFH, the mass-luminosity relation, the age of the disc, the accuracy in stellar distances, the stellar evolutionary models, or the corrections owing to multiple systems, among others, play a significant role. \cite{KTG93}, one of the most referenced works,  derived a slope of $ \alpha= 2.7 \pm 0.4$ explicitly applying  a correction for the unresolved multiple stellar systems mostly for late-type stars. The effects of the unresolved multiple systems on the derivation of the IMF are also discussed in \cite{Sagar1991}, \citealt{Kroupa2003} and, for the high masses, in \cite{Weidner2009}. We want to emphasize that the binary treatment performed here (see Sect. 2.1 and 2.2) enables us to specifically take into account the angular resolution of the catalogues used, thus the treatment of the unresolved systems is rigorous and its effects are implicitly accounted for. More in favour of our IMF at intermediate masses is its consistency with the observed stellar density at Sun position and with the Galactic rotation curve of \cite{Caldwell1981}, all fitted inside the BGM in a consistent scenario that incorporates dynamical constraints (see Sect. \ref{mv}).

To finalize, we cite  the recent work of \cite{Rybizki2015}. These authors, also using  a population synthesis model as a tool for the derivation of the IMF, obtained a slope of $\alpha=3.02$ for supersolar masses, which is only slightly flatter than our value. The strength of their method is their combined use of N-body simulations, Galaxia code (based on BGM as default, see \citealt{Sharma2011}), and Markov chain Monte Carlo techniques to explore the full parameter space. To conclude,  \cite{Haywood1997}, \cite{Rybizki2015} and the present work point towards a slope of the field stars IMF of about $\alpha \approx 3$ at intermediate masses.

\section{Conclusions}\label{conclusions}

Three different statistical methods have been used to search for the best IMF that simultaneously fits both the Galactic classical Cepheids and the whole sky Tycho-2 data. All methods are in agreement with the Kroupa-Haywood IMF (Table 3) with a slope of $\alpha=3.2$ for intermediate masses. Using both field stars and Galactic classical Cepheids, we have found an IMF that is steeper than the canonical stellar IMF for the intermediate masses,  both in associations and young clusters. This result is consistent with the predictions of the Integrated Galactic IMF (IGIMF). The three statistical methods considered here show that a constant SFH is not in agreement with the observational data, thus supporting the star formation history as  decreasing in time in the Galactic thin disc.

For the first time, we use the BGM  to characterise the young population of classical Cepheids and use the most updated boundaries of the Instability Strip. The BGM enables us to properly place the stellar evolutionary models in the context of the Milky Way evolution modelling. The BGM is now capable of providing mass and age distributions of classical Cepheids. We have used the most complete catalogues of Galactic classical Cepheids to confirm these objects as good tracers of the intermediate-mass population when constraining the IMF.  
 
The updated BGM population synthesis model inferred by the Cepheid analysis and undertaken in the present work  represents an improvement on the fit with Tycho-2 data, compared with \cite{Czekaj2014}. With Gaia, thousands of Galactic classical Cepheids will be detected, and the BGM is now ready for the scientific exploitation of these upcoming extremely accurate  data. In a future study, we aim to consider a non-parametric IMF unlinked from any imposed analytical form and use approximate Bayesian computation methods to explore a larger space of parameters using Gaia data.

\begin{acknowledgements}
 This work was supported by the MINECO (Spanish Ministry of Economy) - FEDER through grant ESP2014-55996-C2-1-R and MDM-2014-0369 of ICCUB (Unidad de Excelencia 'Mar\'ia de Maeztu'),  the French Agence Nationale de la Recherche under contract  ANR-2010-BLAN-0508-01OTP and the  European Community's Seventh Framework Programme (FP7/2007-2013) under grant agreement GENIUS FP7 - 606740. We also acknowledge the MareNostrum supercomputer (BSC-CNS) managed by the Barcelona Supercomputing Center.
\end{acknowledgements}

\bibliographystyle{aa}
\bibliography{ref2}

\end{document}